\pdfoutput=1
\documentclass{aa} 

\usepackage[normalem]{ulem}

\usepackage{graphicx}
\usepackage{txfonts}
\usepackage{slashbox}
\usepackage{mhchem}
\usepackage{nccmath}
\usepackage{dblfloatfix}
\usepackage{rotating}
\usepackage{tabularx}
\usepackage[us,12hr]{datetime}
\usepackage{xspace}

\usepackage[normalem]{ulem}
\usepackage[table,xcdraw]{xcolor}

\definecolor{deepmagenta}{rgb}{0.8, 0.0, 0.8}

\usepackage{afterpage}

\usepackage[colorlinks=true,linkcolor=blue,citecolor=blue,urlcolor=blue]{hyperref}

\begin{document}

 \title{Solar oxygen abundance using SST/CRISP center-to-limb observations of the \ion{O}{I} 7772~\AA\ line}

 \author{A.G.M.\ Pietrow\inst{1}\inst{2} 
 \and R.\ Hoppe\inst{3} 
 \and M.\ Bergemann\inst{3} 
 \and F.\ Calvo\inst{2}\inst{4}
     }

 \institute{\inst{1}Leibniz-Institut für Astrophysik Potsdam (AIP), An der Sternwarte 16, 14482 Potsdam, Germany\\
 \inst{2}Institute for Solar Physics, Department of Astronomy, Stockholm University, Albanova University Centre, SE-106 91 Stockholm, Sweden\\
 \inst{3}Max-Planck-Institut für Astronomie, Königstuhl 17, 69117 Heidelberg, Germany\\
 \inst{4}Scientific Computing and Research Support Unit, University of Lausanne, 1015 Lausanne, Switzerland\\
       \email{apietrow@aip.de}\\
      }

\date{Draft: compiled on \today\ at \currenttime~UT}

\abstract{Solar oxygen abundance measurements based on the \ion{O}{I}~near-infrared triplet  have been a much-debated subject for several decades since non-local thermodynamic equilibrium (NLTE) calculations with 3D radiation-hydrodynamics model atmospheres introduced a large change to the 1D LTE modelling. In this work, we aim to test solar line formation across the solar disk using new observations obtained with the SST/CRISP instrument. The observed dataset is based on a spectroscopic mosaic stretching from disk center to the solar limb. By comparing the state-of-the-art 3D NLTE models with the data, we find that the 3D NLTE models provide an excellent description of line formation across the disk. We obtain an abundance value of $A(\mathrm{O}) = (8.73 \pm 0.03)$~dex, with a very small angular dispersion across the disk. We conclude that spectroscopic mosaics are excellent probes for geometric and physical properties of hydrodynamics models and non-LTE line formation.}


 \maketitle
%

\section{Introduction}
Oxygen (O) is the most abundant metal in the universe, and its abundance is an important parameter modern astrophysics and is used widely for determining the metallicity of galaxies \citep[e.g.][]{arellano22}, it influences stellar evolution \citep[e.g.][]{Vandenberg12}, tells us about formation properties and history of exoplanets \citep[e.g.][]{line21} and is an important factor in stellar structure and a major contributor to stellar opacity, and thus an important ingredient for stellar models \citep[e.g.][]{Basu08}. Proper and accurate measurements of the O abundance are thus critical in all of these cases. Traditionally, such methods are tested on the Sun because its disk is spatially resolved and different parts of the atmosphere can be sampled by studying the center-to-limb variation (CLV) of O lines \citep[e.g.][]{Delone74,Dan93,tiago09, bergemann21}. 

\begin{figure}[h]
  \centering
  \includegraphics[width=0.48\textwidth]{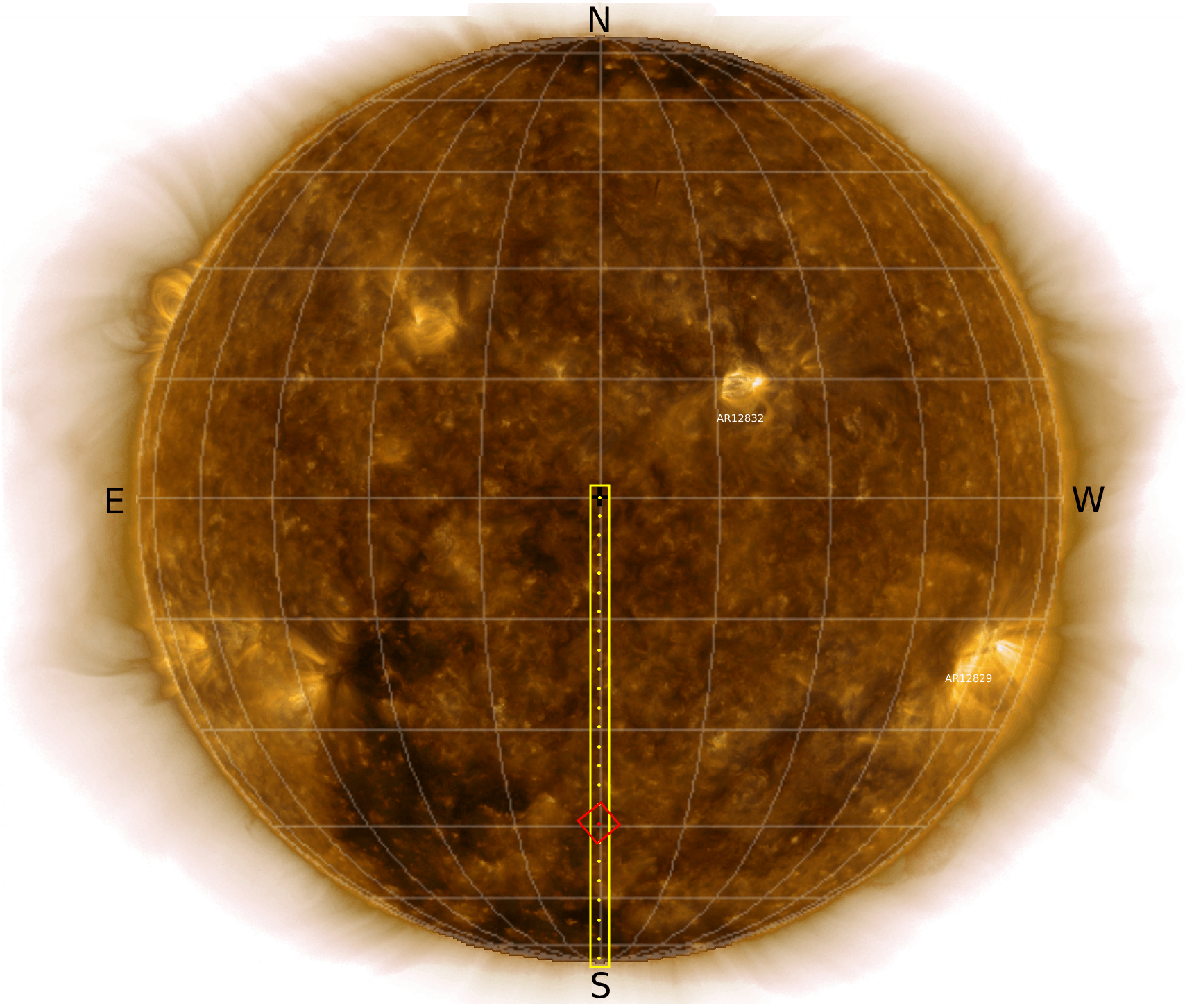}
  \caption{Composite full-disk image of the Sun on 11~June 2021 in the AIA 171, 193 and 304~\AA\ filters \citep{Lemmen2012}. The yellow bar shows 25 pointings of the mosaic which spans from the solar south pole to disk center. The red marker represents a single pointing of the telescope and has the size and orientation of SST/CRISP.}
  \label{fig:pig}
\end{figure}

\begin{figure*}[ht!]
\centering
\includegraphics[width=\textwidth, trim=9cm 0.5cm 7.5cm 1cm, clip]{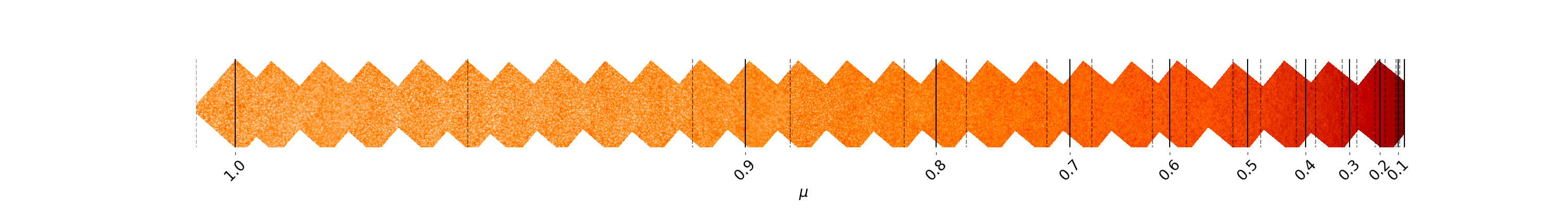}
\caption{Overview of the \ion{O}{I} 7772~\AA\ line: mosaic, starting from disk center on the left and stretching to the solar limb on the right. The black solid lines show 10 positions from $\mu = 0.1$ to 1.0 (cosine of the heliocentric angle) and the dashed lines show bins of $\pm$0.02$\mu$ over which each $\mu$ position was averaged. The mosaics spans roughly $1000\arcsec \times 80\arcsec$.}
\label{fig:clv}
\end{figure*}

The formation of solar O lines has been a subject of interest ever since the (believed) detection of O lines in the solar spectrum \citep{Draper1877, runge1896, Plotkin77}. Photospheric solar abundance studies typically focus on the \ion{O}{I}~6300~\AA~ and the \ion{O}{I}~near-infrared triplet. The former can be modelled in LTE, but suffers from a \ion{Ni}{I} blend that contributes about 25 \% of the equivalent width (EW) of the feature and is sensitive to the treatment of convection \citep{Allende2001,bergemann21}. The latter lines are not significantly affected by blends, but do require to be modeled in NLTE, as was first suggested on an empirical basis by \citet{Magain88} and \citet{spite91} for metal-poor Galactic stars on the grounds of systematic positive difference between the abundances obtained from the  permitted lines and the [O~I] line. Later, the NLTE sensitivity of the 777 nm triplet was confirmed through a detailed theoretical modelling by \citet{Dan91,Dan93} and \citet{dan95} for the Sun.
%
%
%
{In recent years, 3D NLTE modelling} became the norm for the solar photospheric abundance studies. Specifically, for O, the 3D NLTE values include e.g. $(8.76 \pm 0.07)$~dex\footnote{We adopt the traditional astronomical logarithmic abundance scale $A(\epsilon) = 12 + \log_{10}(n_\epsilon / n_\mathrm{H})$, which expresses abundance of element `$\epsilon$' on a logarithmic scale relative to $n_\mathrm{H} = 10^{12}$ hydrogen atoms.} by \citet{caffau08}, $(8.76 \pm 0.02)$~dex by \citet{Steffen15}, $(8.73 \pm 0.05)$~dex by \citet{caffau15},  $(8.69 \pm 0.03)$~dex by \citet{Amarsi18}, $(8.69 \pm 0.04)$~dex by \citet{Asplund21}, $(8.74 \pm 0.03)$~dex by \citet{bergemann21}. Another recent study is that by \citet{magg2022} that employs the O and Ni NLTE model from \citet{bergemann21}, albeit with different spectrum synthesis code and spatially- and temporarily-averaged 3D models similar to \citet{bergemann2012}. Also less model-dependent inference methods based on 3D radiation-hydrodynamics models were used. These methods, for example, were used in \citet{2centro08}, \citet{cubas17}, and \citet{armas20} to derive the solar O abundance of  $A(\mathrm{O}) = 8.86 \pm 0.07$~dex, $8.86 \pm 0.04$~dex, and $8.80 \pm 0.03$~dex, respectively.

%
%
%
%
In an earlier study \citet{bergemann21}, the 3D NLTE oxygen abundance was based on a spectrum with the highest spectral resolution so far, i.e. the solar intensity atlas \citep{Reiners2016} with $R \approx 700,000$ provided by the Institut für Astrophysik Göttingen (IAG). An updated atlas including CLV is currently being prepared for public release \citep{Ellwarth23}. However, it is of an interest to investigate the variation of the solar profile of the \ion{O}{I} lines at a higher spatial resolution across the spectrum because of limited angular sampling of previous spatially-resolved investigations. The aim of this work is to test the consistency of spectral line diagnostics with new SST data and available physical models, and hence help to provide more robust uncertainties on the resulting analysis of photospheric oxygen lines.
%
%
%

We use a new dataset obtained with the CRisp Imaging SpectroPolarimeter \citep[CRISP,][]{Scharmer08} at the Swedish 1-meter Solar Telescope \citep[SST,][]{Scharmer03} as presented in \citet{pietrow22b}. We analyse the CLV of the 7772 \AA~line data published by \citet{pietrow22b} using 1D LTE, 1D NLTE, and 3D NLTE models from \citet{bergemann21} and discuss the implications for the solar O abundance.
\section{Observations and data processing}\label{observations}

The present data were taken as part of a multi-line CLV study \citep{pietrow22b}. We summarize the relevant information below but refer the reader to this paper for a full overview. 

The data consist of a mosaic spanning one solar radius, taken between 10:41 and 11:01 UT on 19 June 2021 with SST/CRISP. The data is reduced using a modified version of the SSTRED pipeline \citep{jaime15,mats21}, which has been designed to process the data from the SST. It not only includes dark and flat-field correction, but typically it also performs image restoration, removing optical aberrations caused by turbulence in the atmosphere (and partially corrected for by the SST adaptive optics) using Multi-Object Multi-Frame Blind Deconvolution \citep[MOMFBD,][]{mats02,vanNoort05}. We omitted this last step, as the reconstruction can fail under poor seeing conditions (Fried-parameter $r_0$ of 5~cm or lower). 

The roughly $60\arcsec \times 1000\arcsec$ mosaic was taken from the solar south pole towards the center (Fig.~\ref{fig:pig}), with roughly 30\% overlap between each consecutive pointing (Fig.~\ref{fig:clv}). The line has been sampled at $\pm$980, $\pm$735, $\pm$392, $\pm$343, $\pm$294, $\pm$245, $\pm$196, $\pm$147,   $\pm$98, $\pm$49, and 0~m\AA\ at $R \approx 160\,000$. We re-calibrate the pixel scale to 0.0584\arcsec~pixel$^{-1}$ by aligning both ends of the mosaic to SDO/HMI \citep{Scherrer2012} observations, which allows us to assign a $\mu$ value at each pixel in the mosaic. The data are then binned into 50 average profiles that are spaced equidistantly in $\mu$ by steps of 0.02. Finally, smoothing the data removes the effects from $p$-modes.

To test the impact of limited resolving power, the data were compared to observations obtained with the Fourier Transform Spectrograph (FTS) at the IAG Vacuum Vertical Telescope (VTT), hereafter referred to as IAG FTS CLV Atlas. The resulting spectra have a resolution of 0.024~cm$^{-1}$ or $R \approx 700\,000$ at $\lambda = 6000$~\AA) \citep[see][]{Reiners2016,schaffer2020,bergemann21}. Afterwards, the HITRAN \citep{hitran} database was used to identify and mask out telluric lines from H$_2$O and O$_2$.

%

\begin{figure}
\centering
\includegraphics[width=1\columnwidth]{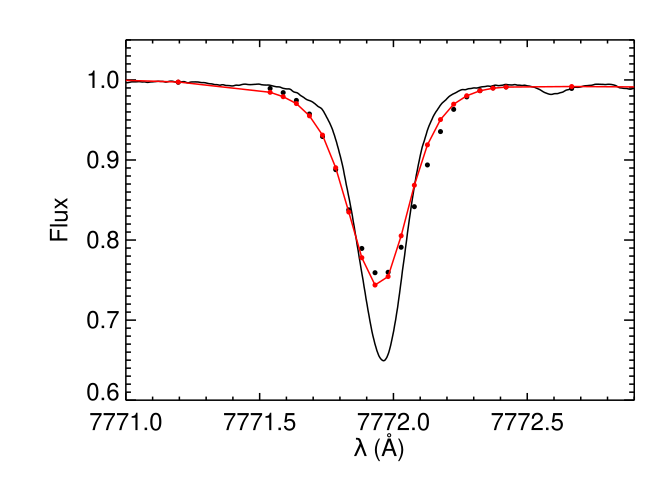}
\caption{Comparison of the observed IAG and SST line profiles for the solar disc centre. The original IAG data are shown with the solid black line. The IAG data degraded to the sampling and resolving power of the SST data are shown with filled black circles. The SST data are shown with filled red circles connected by a red line.}
\label{fig:linprof}
\end{figure}

\section{Methods, results, and discussion}

The abundances of oxygen are calculated using the following approach. For the atmosphere, we used the 3D radiation-hydrodynamics (RHD) simulations of the solar convection from the Stagger grid. We refer to \citet{bergemann2012} and \citet{Magic2013a, Magic2013b} for more details on the RHD models. The NLTE radiation transfer was carried out using the MULTI3D code \citep{Jorrit09}, as updated in \citet{Bergemann2019} and \citet{Gallagher2020}, and the new NLTE model of the oxygen atom developed in \citet{bergemann21}. The model atom furthermore includes new oscillator strengths for the triplet lines from \citet{Bautista2022}. Radiation transfer calculations were carried out using a grid of $80 \times 80 \times 420$ points and corrections for the finite spatial step and the lack of an overlying chromosphere were accounted for in the abundance analysis. 

The line profile analysis follows \citet{bergemann21}, where the abundance was computed via the $\chi^2$ minimization between a series of 3D NLTE line profiles calculated for the chosen $\mu$ values and the observed data. Interpolation between models computed with several values of abundance was applied. To simplify the analysis, we selected 14 $\mu$ positions with a width of $\pm$0.02~$\mu$ from the mosaic that match the observed locations of the IAG data. We estimate the uncertainty of the abundance to be similar to the one given in \citet{bergemann21}, although the resolution of the SST data leads to a slightly larger systematic error. Specifically, the lower sampling of the line and lower resolving power of the SST data makes it harder to correctly describe the line profile (Fig.\ref{fig:linprof}), and the resulting abundance is slightly under-estimated. This was tested by running the analysis on the IAG FTS CLV Atlas degraded to the quality of SST data that can be quantified, e.g. by calculating the line EW. Indeed, the EW of the SST data is somewhat lower compared to the EW of the equivalent IAG profile yielding a $\sim 0.015$ dex difference in the abundance. We note, the EW integration of the SST data is an unreliable procedure, especially because the sampling of the outer wings at $\sim \pm 0.4$ \AA~, where two blends are clearly visible in the original IAG data, is rather poor. A similar systematic  difference is present at other angles across the disc. We account for the systematic bias by folding it into the absolute abundance estimate at each angle. We see a similar offset in Fig. 8 of \citet{pietrow22b}, where the trend matches data from \citet{tiago09}, but a constant shift is found between the two sets.

\begin{figure}[!ht]
\centering
\includegraphics[width=1\columnwidth]{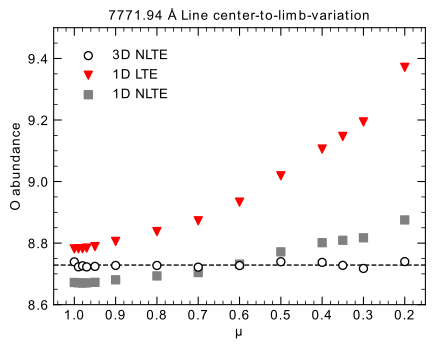}
\caption{Variation of oxygen abundance across the solar disk obtained from the SST data fitted with 3D NLTE models (open circles), a 1D LTE model (closed triangles), and a 1D NLTE model (closed squares).
}
\label{fig:results}
\end{figure}

The resulting oxygen abundances derived from the SST/CRISP data for 14 $\mu$ points across the solar disk are presented in Fig.~\ref{fig:results}. The average abundance obtained from the 3D NLTE modelling is very precise and independent of the viewing angle, yielding $A(\mathrm{O}) = (8.73 \pm 0.031)$~dex. For comparison, we also show the results obtained from the 1D LTE and NLTE modelling. Clearly, the latter models are unable to describe the properties of observed solar radiation field and its CLV, over-estimating the oxygen abundance by more than 0.4~dex at the limb. This supports the previous result from \citep{bergemann21} and reinforces the evidence that the 3D NLTE models are sufficiently complete to provide a realistic description of oxygen line formation across the disk. Hence, the abundance estimates obtained using 3D NLTE models are to be preferred for precision stellar spectroscopic diagnostics. 
%
%
%
%

\section{Conclusions}\label{conclusions}
By comparing the 3D NLTE oxygen models from \citet{bergemann21} with our new spatially resolved SST/CRISP data, we find that the solar oxygen abundance $A(\mathrm{O}) = (8.73 \pm 0.031)$~dex, is fully consistent with the earlier result. Our data do not reveal any angular dependence of abundance, reinforcing the accuracy of the 3D NLTE modeling approach as compared to 1D modeling. The abundance is consistent with the value from \citet{Amarsi18} within the respective uncertainties of both estimates. The difference between their study and our result is indeed rather modest, given the differences in the choice of observations, gf-values, correction for the chromospheric back-heating, as well different 3D NLTE codes and the NLTE model of O. We conclude that the CLV datasets from \citet{pietrow22b} and those of the IAG FTS CLV Atlas complement each other and synergistically probe the geometric and physical properties of RHD models of stellar convection and non-LTE line formation. However, higher resolution spectral data are preferred where possible for precision solar abundance diagnostics.


\begin{acknowledgements}
AP was supported at AIP by grants the European Commission’s Horizon 2020 Program under grant agreements 824064 (ESCAPE -- European Science Cluster of Astronomy \& Particle Physics ESFRI Research Infrastructures) and 824135 (SOLARNET -- Integrating High Resolution Solar Physics). The CHROMATIC project (2016.0019) of the Knut and Alice Wallenberg foundation supported AP at SU.
MB is supported through the Lise Meitner grant from the Max-Planck Gesellschaft. We acknowledge support by the Collaborative Research center SFB 881 at Universität Heidelberg (projects A5, A10) of the Deutsche Forschungsgemeinschaft (DFG). MB received funding from the European Research Council (ERC) as part of the European Commission’s Horizon 2020 program under grant agreement 949173.
We thank Carsten Denker for his feedback on the manuscript text and the anonymous referee for their valuable suggestions during the peer-review process.
The Swedish 1-meter Solar Telescope is operated on the island of La Palma by the Institute for Solar Physics of Stockholm University in the Spanish Observatorio del Roque de los Muchachos of the Instituto de Astrof\'isica de Canarias. The Institute for Solar Physics was supported by a grant for research infrastructures of national importance from the Swedish Research Council (registration number 2017-00625).

This research has made use of NASA's Astrophysics Data System (ADS) bibliographic services. 
We acknowledge the community efforts devoted to the development of the following open-source packages that were used in this work: numpy (\href{http:\\numpy.org}{numpy.org}), matplotlib (\href{http:\\matplotlib.org}{matplotlib.org}), and astropy (\href{http:\\astropy.org}{astropy.org}).
We extensively used the CRISPEX analysis tool \citep{Gregal12}, the ISPy library \citep{ISPy2021}, and SOAImage DS9 \citep{2003DS9} for data visualization.

\end{acknowledgements}

\bibliographystyle{aa}
\bibliography{ref}

\end{document}